\documentclass{PoS}

\usepackage{graphicx}

\DeclareMathAlphabet{\mathpzc}{OT1}{pzc}{m}{it}

\usepackage{multirow}

\usepackage{amsmath}
\usepackage{nicefrac}
\usepackage{amssymb}
\usepackage{upgreek}
\usepackage{esint}
\usepackage{slashed}

\def\beq{\begin{equation}}
\def\eeq{\end{equation}}
\def\bea{\begin{eqnarray}}
\def\eea{\end{eqnarray}}
\def\beqa{\begin{equation}\begin{array}{l}}
\def\eeqa{\end{array}\end{equation}}
\def\eqlab#1{\label{eq:#1}}
\def\figlab#1{\label{fig:#1}}

\def\eref#1{(\ref{eq:#1})}
\def\Eqref#1{Eq.~(\ref{eq:#1})}

\def\Figref#1{Fig.~\ref{fig:#1}}

\def\barr{\left(\begin{array}{c}}
\def\earr{\end{array}\right)}
\def\bmat{\left(\begin{array}{cc}}
\def\emat{\end{array}\right)}
\def\al{\alpha}
\def\be{\beta}
\def\ga{\gamma} 
 
\def\veps{\varepsilon}  

\def\la{\lambda} \def\La{{\Lambda}}

\def\si{\sigma}

\def\nn{\nonumber}
\def\dd{\mathrm{d}}

\def\cA{{\mathcal A}}

\DeclareMathOperator\arccosh{arccosh}

\title{Pseudoscalar-Meson Contributions to $g-2$ \\ via Schwinger's Sum Rule}

\ShortTitle{Pseudoscalar-Meson Contributions to $g-2$ via Schwinger's Sum Rule}

\author{\speaker{Franziska Hagelstein}\\
        Albert Einstein Center for Fundamental Physics, Institute for Theoretical Physics, \\University of Bern, Sidlerstrasse 5, CH-3012 Bern, Switzerland\\
        E-mail: \email{hagelstein@itp.unibe.ch}}
        
        \author{Vladimir Pascalutsa\\
        Institut f\"ur Kernphysik \& Cluster of Excellence PRISMA,  \\
        Johannes Gutenberg-Universit\"at Mainz, D-55128 Mainz, Germany\\
        E-mail: \email{pascalut@uni-mainz.de}}

\abstract{The Schwinger sum rule is presented as a new promising tool to study the hadronic contributions to the muon anomalous magnetic moment. In particular, we show preliminary results for the light-by-light scattering contribution of pseudoscalar mesons.}

\FullConference{The 9th International workshop on Chiral Dynamics\\
		17-21 September 2018\\
		Durham, NC, USA}

\begin{document}

\section{Introduction}
\noindent Presently, there is about $4\,\sigma$ discrepancy between the experimental value of the muon anomalous magnetic moment (AMM), $a_\mu=\nicefrac{1}{2}(g-2)_\mu$, from BNL-E821 \cite{Bennett:2006fi} and the Standard Model (SM) prediction \cite{Jegerlehner:2018zrj}:
\begin{subequations}
\bea
a_\mu^\text{exp.}&=&[11 \, 659\, 209.1 \pm6.3]\times 10^{-10},\\
a_\mu^\text{th.}&=&[11\, 659\, 178.3 \pm4.3]\times 10^{-10},
\eea
\end{subequations}
where the theory uncertainty
 is dominated by the hadronic vacuum polarization (HVP) and hadronic light-by-light scattering (HLbL) contributions, see \Figref{Hamm} (a) and (b), respectively:
\begin{subequations}
\bea
a_\mu^\text{LO HVP}&=&[689.46 \pm 3.25]\times 10^{-10}, \qquad \text{\cite{Jegerlehner:2017zsb}}\eqlab{LOHVPnum}\\
a_\mu^\text{HLbL}&=&[10.34\pm 2.88]\times 10^{-10}. \qquad \text{\cite{Jegerlehner:2017gek}}\eqlab{totalhLbL}
\eea
\end{subequations}
The ongoing experiments at Fermilab \cite{Logashenko_2015,Venanzoni:2014ixa} and J-PARC \cite{Otani:2015jra} are expected to improve the experimental precision by a factor of $4$, as well as to provide an important cross check of the previous experiment and of each other. The SM prediction should receive a complementary improvement, meaning 
the uncertainty of the hadronic contributions should  reduce substantially.

The HVP contribution admits a simple dispersive formula:
\beq
\eqlab{HVPdisp}
a_\mu^\text{LO HVP} = \frac{1}{4\pi^3}\int_{s_0}^\infty \dd s\, \sigma^{\,e^+e^- \rightarrow\; \mathrm{hadrons}}(s)
\int_0^1 \dd x \,\frac{x^2(1-x)}{x^2 + (1-x)(s/m_\mu^2)} \, ,
\eeq
which, 
to leading order in the fine structure constant $\al$, determines it through a single observable: the total cross section of $e^+e^- \to \mbox{hadrons}$.
The common treatment of the HLbL contribution is much more involved and model-dependent
(cf., Ref.~\cite{Jegerlehner:2017gek}), which is basically why the relative accuracy
of the HLbL value is so much worse than that of the HVP. It is certainly desirable to 
have an analogue of the simple formula \eref{HVPdisp} for the HLbL contribution,
and that brings us to the subject of this talk.

\begin{figure}[bh] 
    \centering 
       \includegraphics[height=2.5cm]{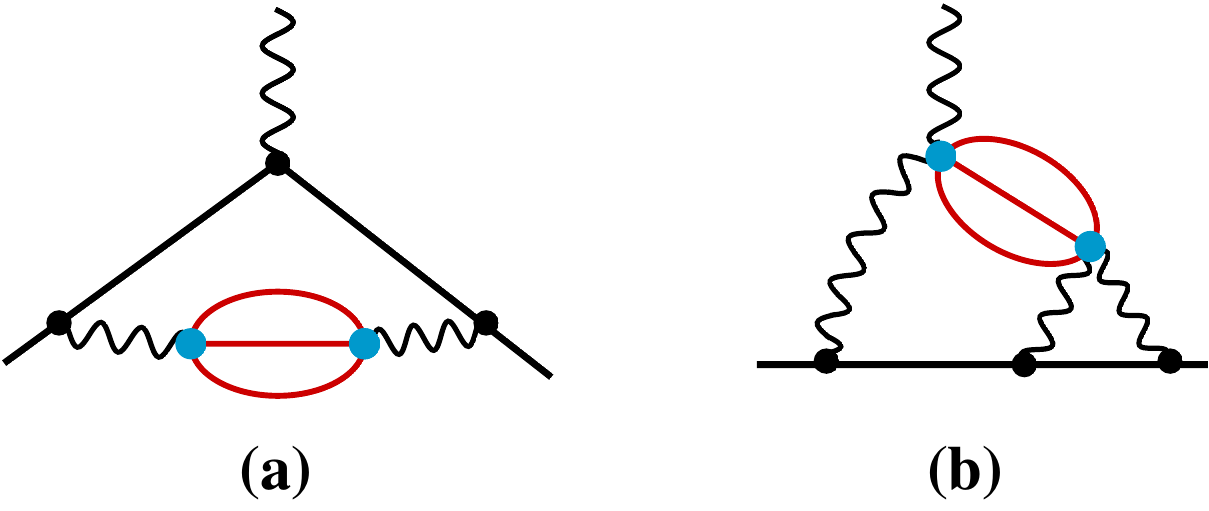}
\caption{Hadronic contributions to the anomalous magnetic moment: (a) hadronic vacuum polarization and (b) hadronic light-by-light scattering. Hadronic excitations are indicated by red blobs. \figlab{Hamm}}
\end{figure}

\section{The Schwinger Sum Rule}

As we recently argued \cite{Hagelstein:2017obr}, the Schwinger sum rule encompasses
the dispersive formula \eref{HVPdisp} for HVP and provides its analogue for HLbL (or,
in fact,
any other contribution). 
The Schwinger sum rule reads \cite{Schwinger:1975ti,HarunarRashid:1976qz}: 
\begin{subequations}
\eqlab{newSR}
\bea
a_\mu & =& \frac{m_\mu^2}{\pi^2\alpha}
\int_{\nu_0}^\infty \!\dd \nu \, \left[\frac{\si_{LT}(\nu,Q^2)}{Q}\right]_{Q^2=0}\\
&=&  \lim_{Q^2\to 0} \frac{8m_\mu^2}{Q^2}
 \int_0^{x_0} \dd x \;  
 \big[ g_1(x,Q^2) + g_2(x,Q^2)\big] ,
 \eqlab{SSR2}
\eea
\end{subequations}
where $\sigma_{LT}(\nu,Q^2)$ is a doubly-polarized photoabsorption cross section, 
for a given energy $\nu$ and virtuality $Q^2$ of the photon. The polarization
of the absorbed photon and the target (here, the muon) is rather peculiar in this
observable: for the photon it is the interference of the
longitudinal and transverse polarization, whereas for the target it is the interference
between the positive and negative helicity. One can also understand
this observable in terms of the standard spin structure functions, $g_1$ and $g_2$, 
see \Eqref{SSR2}, with the Bjorken variable $x=Q^2/(2m_\mu \nu)$.

To see how this sum rule works in QED, recall that the leading photoabsorption
process therein is Compton scattering, described by the Feynman diagrams in \Figref{treeCS}.
The corresponding helicity amplitudes are
given by:
\beq 
\mathcal{M}_{\la'h'\la h} = -e^2 \veps_{\la'}^{\ast\,\mu} (q') \veps_{\la}^\nu (q)
\, \bar{\mathpzc{u}}_{h'} (p') \left[ \gamma_\mu \,\frac{\slashed{p}+\slashed{q}+m_\mu}{s-m_\mu^2}\,\ga_\nu+\ga_\nu\, \frac{\slashed{p}'-\slashed{q}+m_\mu}{u-m_\mu^2}\,\gamma_\mu\right]\mathpzc{u}_{h}(p),
\eeq 
where $\veps(q)$ and $\la$ denote the photon polarization-vector and helicity, 
whereas $\mathpzc{u}(p)$ and $h$ stand for the lepton spinor and helicity; 
the Mandelstam variables are qiven as usual by $s=(p+q)^2$, $u=(p'-q)^2$, $t=(q-q')^2$; $e^2=4\pi\al$.
The cross section entering the Schwinger sum rule is then defined via:
\beq 
\frac{\dd \si_{LT}}{\dd t} =
\frac{1}{16\pi (s-m_\mu^2)^2} \sum_{\la' h'} \frac{1}{\sqrt{2}} \mathrm{Re}\,\big( \mathcal{M}^\ast_{\la'h'\,1\, \nicefrac{1}{2}}
\mathcal{M}_{\la'h'\, 0 \, -\nicefrac{1}{2}}\big) .
\eqlab{diffCS}
\eeq

Upon integrating over $t$, we obtain the total $LT$ cross section for the tree-level (virtual) Compton scattering:
\bea  
 \si_{LT}(\nu, Q^2) =\frac{\pi  \alpha ^2 Q \,(s-m_\mu^2)^2 }{4 m_\mu^3 \nu ^2 \big(\nu ^2+Q^2\big)} \left( -2 - \frac{m_\mu(m_\mu+\nu)}{s}
+ \frac{3 m_\mu+2 \nu}{\sqrt{\nu ^2+Q^2}}
   \, \mathrm{arccoth} \frac{m_\mu+\nu}{\sqrt{\nu ^2+Q^2}}\right),
\eea
with $\nu = p\cdot q/m_\mu = (s-m_\mu^2+Q^2)/2m_\mu$, $Q^2=-q^2$. Substituting this expression into the Schwinger sum rule, we
reproduce another famous result of Julian Schwinger: $\al/2\pi$. 

\begin{figure}[h]
\centering
\includegraphics[width=0.4\textwidth]{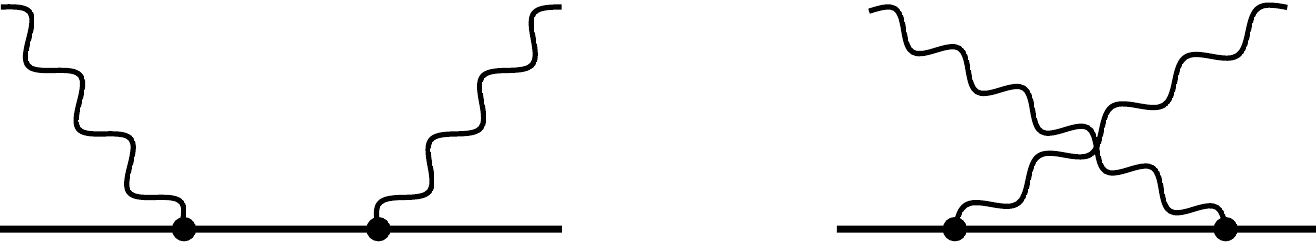}
\caption{Tree-level Compton scattering diagrams.}
\figlab{treeCS}
\end{figure}

\section{Unified Treatment of Hadronic Contributions} \label{channelsSection}

To evaluate the impact of a given mechanism on $a_\mu$ via the Schwinger sum rule,  we need to
measure its effect on the  photoabsorption cross section $\si_{LT}$. There are two fundamentally
different ways in which hadrons affect the photoabsorption on a lepton:
\begin{itemize}
    \item[(i)] hadron photoproduction, e.g., \Figref{Hphotoprod} (a) and (b);
     \item[(ii)] hadronic effects in the electromagnetic channels, e.g., \Figref{Hphotoprod} (c) and (d).
\end{itemize}
\begin{figure}[h]
\centering
\includegraphics[height=2.5cm]{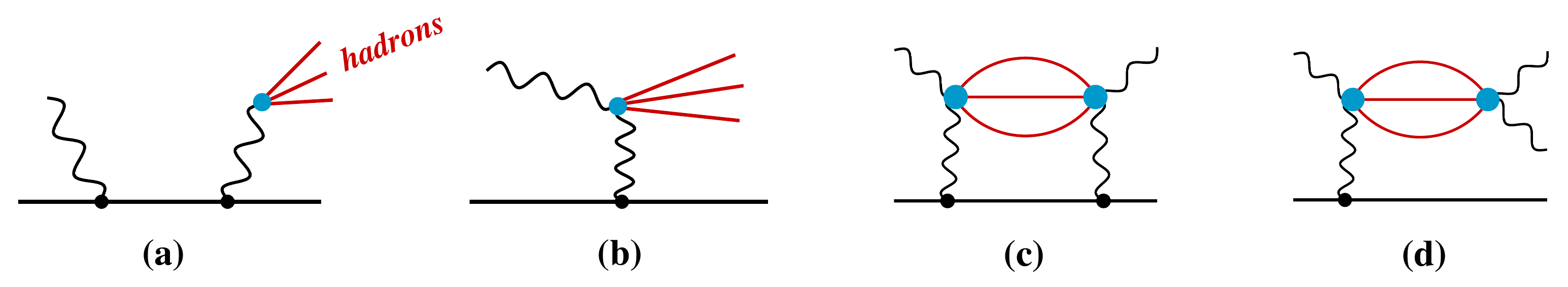}
\caption{Different channels contributing to the photoabsorption process: hadron photoproduction through timelike Compton scattering (a) or the Primakoff mechanism (b), and hadronic light-by-light contributions to the Compton scattering (c, d)  (crossed diagrams omitted).}
\figlab{Hphotoprod}
\end{figure}
The first type of contributions can in principle be measured experimentally. 
They begin to contribute to $\sigma_{LT}$ at $\mathcal{O}(\al^3)$,
and hence are of  $\mathcal{O}(\al^2)$ in $a_\mu$. As shown in Ref.~\cite{Hagelstein:2017obr}, 
the mechanism of \Figref{Hphotoprod} (a) (timelike Compton scattering) by itself
yields exactly the leading-order HVP formula \eref{HVPdisp}. The mechanism of \Figref{Hphotoprod} (b)
(Primakoff) by itself yields a vanishing contribution to $a_\mu$ \cite{Hagelstein2019}, which can be proven exactly using the
sum rules for  light-by-light scattering (see, e.g.,~\cite{Pascalutsa:2012pr}). 
The Primakoff mechanism may contribute in the interference
with subleading effects, as will be considered below for the pseudoscalar-meson production, see \Figref{Figure3_PionProduction}.

In evaluating the hadronic
effects of type (ii), which to $\mathcal{O}(\al^3)$ in $a_\mu$ are given
by the interference of the diagrams in \Figref{Hphotoprod} (c, d) with the tree-level QED diagrams, 
one faces the same sort of problem as in the evaluation of the
entire HLbL contribution \Figref{Hamm}(b), albeit at less than two-loop level. The reduction in number
of loops provides a significant simplification and 
should be helpful in a better determination of the total HLbL contribution.

\section{Pseudoscalar-Meson Contribution}

\begin{figure}[b]
    \centering 
\includegraphics[height=2cm]{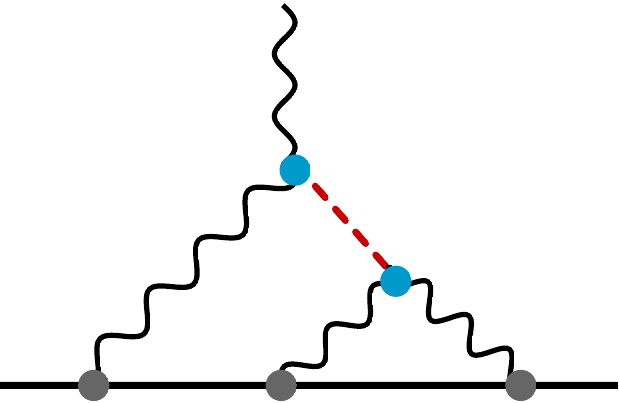}
\caption{Single-meson (red dashed line) contribution to the anomalous magnetic moment.\figlab{HLbLpi0}}
       \end{figure}

\noindent  The neutral pseudoscalar mesons $\pi^0$, $\eta$ and $\eta'$ play a significant role in the HLbL contribution, see \Figref{HLbLpi0}. The model evaluations of the so-called ``pseudoscalar-pole'' contribution may differ depending on what is used for the meson transition form factors (or, whether they should be used in both vertices or just one). The most commonly quoted values are:
\begin{subequations}
\eqlab{PSpole}
\bea
a_\mu^{\text{PS-pole}}&=&8.3\pm1.2\times 10^{-10}\qquad\text{Knech \& Nyffeler \cite{Knecht:2001qf},}\eqlab{KNvalue}\\
a_\mu^{\text{PS-pole}}&=&11.4\pm1.0\times 10^{-10}\qquad\text{Melnikov \& Vainshtein \cite{Melnikov:2003xd}.}\eqlab{MVvalue}
\eea
\end{subequations}
The most recent value for the dominant $\pi^0$-pole contribution \cite{Hoferichter:2018dmo}:
\beq
a_\mu^{\pi^0\text{-pole}}=6.26\,^{+0.30}_{-0.25}\times 10^{-10}, \eqlab{pi0Hoferichter}
\eeq
concurs with the first of the two calculations in \Eqref{PSpole}.

It is interesting to see whether the Schwinger sum-rule approach can tell us something new.
To leading order in $\al$, we need to consider the following 
two hadronic channels: 
$\gamma \mu  \rightarrow \mu \pi^0 $ and  $\gamma\mu  \rightarrow \mu  \pi^0 \gamma$, for each of pseudoscalars.
Here, we present results for the calculation of the first channel:  $\gamma \mu  \rightarrow \mu \,(\pi^0, \eta, \eta') $, see \Figref{Figure3_PionProduction}. The same calculation  will apply to the electron AMM, $a_e$.
\begin{figure}[t]
\centering
\includegraphics[width=0.6\textwidth]{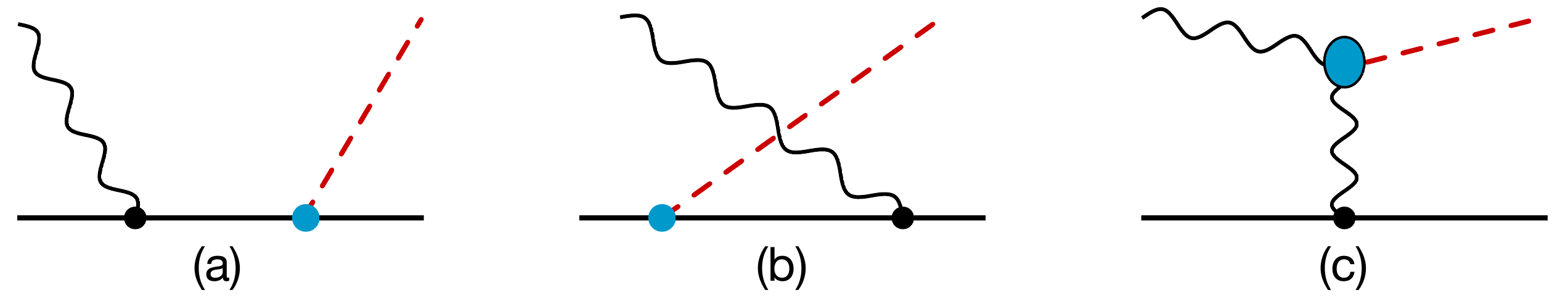}
\caption{Diagrams contributing to the process of pseudoscalar-meson  production off the muon. }
\figlab{Figure3_PionProduction}
\end{figure}

In order to compute the diagrams in \Figref{Figure3_PionProduction}, we 
first of all need to specify how a pseudoscalar meson couples to photons and leptons. We use the following couplings:
\begin{subequations}
\bea
\Gamma^{\mu \nu}_{\pi \ga \ga}(q,q')&=&-ie^2 F_{\pi \ga^* \ga^*}(q^2,q^{\prime\,2})\,\epsilon^{\mu \nu \al \beta} q_{\al} q'_{ \beta},\\
\Gamma_{\pi \ell \ell}&=&-F_{\pi \ell \ell}(M_\pi^2,m^2,m^2)\, \ga_5,\eqlab{VertexPill}
\eea
\end{subequations}
each characterized by a form factor. For the coupling to photons, it is the standard transition form factor. Here, the momentum $q$ and the index $\nu$ describe the incoming photon, $q'$ and $\mu$ the outgoing photon.
\begin{figure}[b]
\centering
\includegraphics[width=0.6\textwidth]{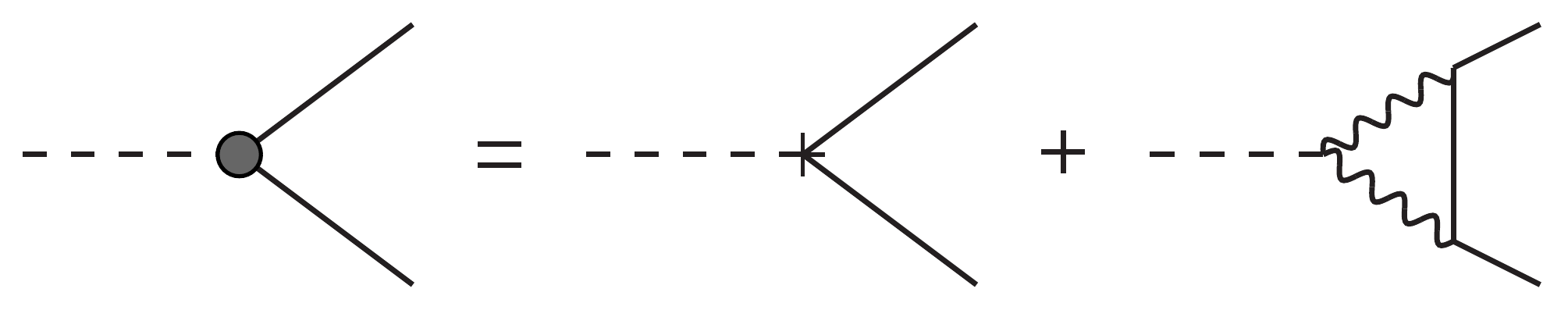}
\caption{Leading contributions to the pseudoscalar-lepton-lepton interaction.}
\figlab{VertexPi0}
\end{figure}
For the $\pi \ell \ell$ form factor (where $\ell$ stands for a lepton with mass $m$ and $\pi$ for a pseudoscalar meson with respective mass $M_\pi$),
we use the well-known dispersion relation \cite{Drell59,Ametller:1984uk} (resulting from the  diagrams in \Figref{VertexPi0}):
\begin{subequations}
\eqlab{muPiFF}
\bea
 F_{\pi \ell \ell}(q^2)& \equiv & F_{\pi \ell \ell}(q^2,m^2,m^2)  \eqlab{Pion0FF}= F_{\pi \ell \ell}(0) + \frac{q^2}{\pi} \int_0^\infty \!\dd s\,  \frac{\mathrm{Im}\, F_{\pi \ell \ell}(s)}{(s-q^2) s}  , \\
 \mathrm{Im}\, F_{\pi \ell \ell}(s) &=& -\frac{\al^2 m}{2\pi f_{\pi}} \frac{\arccosh(\sqrt{s}/2m)}{\sqrt{1-4m^2/s}}, \\
 F_{\pi \ell \ell}(0)  &=&  \frac{\al^2 m}{2\pi^2 f_{\pi}} \Big[ \cA (\La) + 3\ln\frac{m}{\La} \Big],\eqlab{muPiFFc}
\eea 
\end{subequations}
where $\La$ is the renormalization scale, and $\cA$ is a pion-lepton low-energy constant. We evaluate $\cA$ analogously to Ref.~\cite[Eq.~(10) and (11)]{Dorokhov:2007bd}, where we use the LMD+V model for the pion transition form factor \cite[Eq.~(4.4)]{Knecht:2001qf} and VMD models for the eta and eta-prime transition form factors \cite[Eq.~(37)]{Nyffeler:2009tw}. In addition, we make use of the 
experimentally known decay widths \cite{Tanabashi:2018oca}:
\begin{subequations}
\eqlab{widths}
\bea
\Gamma(\pi^0\to e^+ e^-)&=&0.499(28)\,\upmu\text{eV},\\
\Gamma(\eta\to \mu^+ \mu^-)&=&7.6(1.1)\,\text{meV},
\eea
\end{subequations}
which are related to the $\pi \ell \ell$ form factor in the following way \cite{Griffiths}:
\beq
\hspace{-0.2cm}\Gamma(\pi\to \ell^+ \ell^-) = \frac{M_\pi}{8\pi} \sqrt{1-\frac{4m^2}{M_\pi^2}} \, \big| F_{\pi \ell \ell}(M_\pi^2,m^2,m^2) \big|^2,
\eeq
and analogously for the other leptonic decays of pseudoscalar mesons. The resulting couplings are listed in Table \ref{Tab1Coupling}, where the errors stem from the decay widths in \Eqref{widths}.

\renewcommand{\arraystretch}{1}
\begin{table}[t]
\caption{Coupling strengths of pseudoscalar-muon-muon interactions.  \label{Tab1Coupling}}
\begin{center}
\begin{tabular}{ |c|c|c|c| }
\hline
&$\pi^0$&$\eta$&$\eta'$\\
\hline

$\cA(m)$&$-5.4^{\,+1.0}_{\,-1.2}$&$-6.9_{\,-0.8}^{\,+1.1}$&$-6.0$\\
$\mathrm{Re}\left[F_{\pi \ell \ell}(M_\pi^2,m^2,m^2)\right]\times 10^7$&$-145^{\,+32}_{\,-36}$&$-98_{\,-24}^{\,+33}$&$-11$\\
 \hline
\end{tabular}
\end{center}
\end{table}
\renewcommand{\arraystretch}{1}

 The  interference between the diagrams \Figref{Figure3_PionProduction} (a, b) and \Figref{Figure3_PionProduction} (c), needed to evaluate the $\mathcal{O}(\al^3)$ contribution to $a_\mu$, 
 gives the following expression (omitting helicities):
\bea
\eqlab{Msquared}
\vert \mathcal{M}\vert^2_{\mathrm{interf.}}&=& ie^4\, \mathrm{Re}\left[F_{\pi \ell \ell}(M_\pi^2,m_\mu^2,m_\mu^2)\right]\,\frac{F_{\pi \ga^* \ga^*}(-Q^2,t)}{t}\,\epsilon^{\mu \nu \al \be}q_\al \left(p'-p\right)_\be \\
&\times &\left\{\veps^*_\nu(q)\;\bar{\mathpzc{u}}(p) \gamma_\mu \mathpzc{u}(p') \;\bar{\mathpzc{u}}(p')\left[ \gamma_5 \,\frac{\slashed{p}+\slashed{q}+m_\mu}{s-m_\mu^2}\,\slashed{\veps}(q)+\slashed{\veps}(q)\, \frac{\slashed{p}'-\slashed{q}+m_\mu}{u-m_\mu^2}\,\gamma_5\right]\mathpzc{u}(p)\right.\nn\\
&+&\left.\bar{\mathpzc{u}}(p)\left[ \slashed{\veps}^*(q) \,\frac{\slashed{p}+\slashed{q}+m_\mu}{s-m_\mu^2}\,\gamma_5+\gamma_5\, \frac{\slashed{p'}-\slashed{q}+m_\mu}{u-m_\mu^2}\,\slashed{\veps}^*(q)\right]\mathpzc{u}(p')\; \bar{\mathpzc{u}}(p') \gamma_\mu \mathpzc{u}(p) \;\veps_\nu(q)\right\},\nn
\eea
where $p$ ($p'$) is the initial (final) lepton momentum,  $q$ is the incoming photon momentum, and the Mandelstam variables are defined as usual.
Substituting this into \Eqref{diffCS}, we obtain the differential cross section, which in the
limit of $Q^2\to 0$ takes the following form:
\bea
\left[\frac{1}{Q}\frac{\dd \sigma_{LT}}{\dd t}\right]_{Q^2=0}&=&\frac{\al^2 \pi}{8m_\mu^2} \,F_{\pi\,\gamma^* \gamma^*}(t,0) \, \mathrm{Re}\left[F_{\pi \ell \ell}(M_\pi^2,m_\mu^2,m_\mu^2)\right]
\frac{1}{ t (s-m_\mu^2)(u-m_\mu^2)\nu^4}\\
&\times&\Big\{m_\mu^2 M_\pi^4 \left(M_\pi^2-2 \nu ^2\right)-t^3\,m_\mu\left(m_\mu+2 \nu\right)\nn\\
&&\quad-\; t  \left[m_\mu^2 \left(8 \nu ^4+3 M_\pi^4-8 \nu ^2 M_\pi^2\right)+2\nu\, m_\mu   M_\pi^2 \left(2 \nu ^2+M_\pi^2\right)-2 \nu ^2 M_\pi^4\right]\nn\\
&&\quad+\;t^2\left[3 m_\mu^2 \left(M_\pi^2-2 \nu ^2\right)+4 \nu  m_\mu \left(M_\pi^2-\nu ^2\right)-2 \nu ^2 M_\pi^2\right]\Big\}.\nn
\eea
To obtain the integrated cross section, we integrate within the following interval:
\eqlab{tBoundaries}
\beq
t_{\substack{\text{max}\\ \text{min}}}= M_\pi^2-(s-m_\mu^2) \big[ \beta(s)\mp \la(s)\big],
\eeq
with $\beta(s) = (s-m_\mu^2+M_\pi^2)/2s$, $\,\la(s)=(1/2s)\sqrt{[s-(M_\pi+m_\mu)^2][s-(M_\pi-m_\mu)^2]}$.

We are now left to evaluate the sum rule integral, \Eqref{newSR}, with the photoabsorption cross section derived above, starting from the pseudoscalar-production threshold ($Q^2=0$):
\beq
\nu_0=M_\pi\left(1+\frac{M_\pi}{2m_\mu}\right).
\eeq
Our results for the contribution of the pseudoscalar-meson production channel to $a_\mu$, derived via the Schwinger sum rule, are shown in Table \ref{Tab0}. The errors for the $\pi^0$ and $\eta$ channels are propagated from the pseudoscalar-muon-muon couplings in Table \ref{Tab1Coupling}. For the contribution of the $\eta'$ production, we assigned a $33\%$ error, similar to the maximal error found for the $\eta$ production. Note that in our calculation we are neglecting the effect of off-shell muons in \Figref{Figure3_PionProduction} (a, b), as well as the diagram in \Figref{OffShellEffect}.
\begin{figure}[h]
\centering
\includegraphics[width=0.15\textwidth]{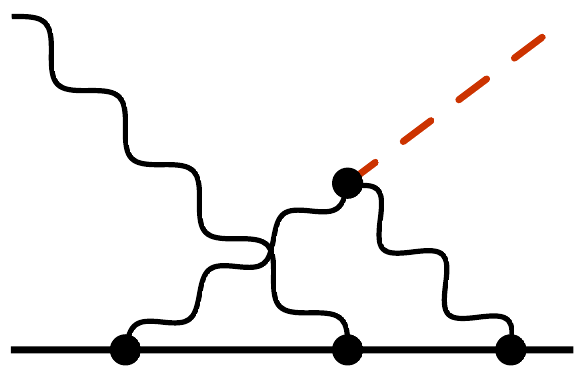}
\caption{Subleading mechanism accompanying the single-meson photoproduction.}
\figlab{OffShellEffect}
\end{figure}

\renewcommand{\arraystretch}{1}
\begin{table}[t]
\caption{Contribution of the pseudoscalar-meson production channel to $a_\mu$ in units of $10^{-10}$. \label{Tab0}}
\begin{center}
\begin{tabular}{ |c|c|c|c| }
\hline
$\ga \ell \rightarrow \ell\pi^0$&$\ga \ell \rightarrow \ell\eta$&$\ga \ell \rightarrow \ell\eta'$&$\gamma \ell  \rightarrow \ell \,(\pi^0, \eta, \eta') $\\
\hline
$16.6^{\,+4.1}_{\,-3.7}$&$8.5^{\,+2.1}_{\,-2.9}$&$1.0\pm0.3$&$26.1^{\,+4.6}_{\,-4.7}$\\
\hline
\end{tabular}
\end{center}
\end{table}
\renewcommand{\arraystretch}{1}

\section{Conclusion}
The Schwinger sum rule provides a dispersive data-driven approach to calculating the hadronic contributions in lepton  $g-2$. 
It encompasses the simple dispersive formula for the HVP contribution, \Eqref{HVPdisp}, and allows to treat
the others (e.g, the HLbL contribution) on similar footing. The required data on the doubly-polarized photoabsorption
cross section $\sigma_{LT}$, which should be used in the sum rule as input, are not presently available. 
In the absence of data, we are setting up a model for the hadron photoproduction process on the muon.

We have so far evaluated the contribution of the pseudoscalar-meson production channels ($\gamma \mu  \rightarrow \mu \,\pi^0, \mu\eta, \mu\eta' $)  to the muon $g-2$. More specifically, we have calculated the interference cross section between the diagrams in \Figref{Figure3_PionProduction} (a, b) with the Primakoff diagram in  \Figref{Figure3_PionProduction}  (c).
The resulting contribution to the muon $g-2$, given in Table \ref{Tab0},   is a factor $2$ to $3$ larger than the conventional pseudoscalar-pole contributions of Eq.~\eref{PSpole} and \eref{pi0Hoferichter}. 
However, as explained in Sec.~\ref{channelsSection}, the single-meson photoproduction channel  is one of four photoabsorption channels contributing to $g-2$ at $\mathcal{O}(\al^3)$.  In other words, our results are only a partial $\mathcal{O}(\al^3)$ calculation of the HLbL contribution, and hence, may not be 
directly comparable to the conventional meson-pole results. 
Inclusion of the remaining channels might restore the agreement with (at least one of) the conventional calculations.

\section*{Acknowledgements}
\noindent This work was supported by the Swiss National Science Foundation, and the Deutsche Forschungsgemeinschaft (DFG) 
through the Collaborative Research Center [The Low-Energy Frontier of the Standard Model (SFB 1044)].


\begin{thebibliography}{99}
 
\bibitem{Bennett:2006fi} 
  G.~W.~Bennett, {\it et al.} [Muon g-2 Collaboration],
  Phys.\ Rev.\ D {\bf 73}, 072003 (2006).
  
\bibitem{Jegerlehner:2018zrj} 
  F.~Jegerlehner,
  Acta Phys.\ Polon.\ B {\bf 49}, 1157 (2018).
  
\bibitem{Jegerlehner:2017zsb} 
  F.~Jegerlehner,
  arXiv:1711.06089 [hep-ph].
 
\bibitem{Jegerlehner:2017gek} 
  F.~Jegerlehner,
  Springer Tracts Mod.\ Phys.\  {\bf 274} (2017).
  
\bibitem{Logashenko_2015} 
  I.~Logashenko, {\it et al.} [Muon g-2 Collaboration],
  J.\ Phys.\ Chem.\ Ref.\ Data {\bf 44}, no. 3, 031211 (2015).
  
\bibitem{Venanzoni:2014ixa} 
  G.~Venanzoni [Fermilab E989 Collaboration],
  Nucl.\ Part.\ Phys.\ Proc.\  {\bf 273-275}, 584 (2016).
  
\bibitem{Otani:2015jra} 
  M.~Otani [E34 Collaboration],
  JPS Conf.\ Proc.\  {\bf 8}, 025008 (2015).

\bibitem{Hagelstein:2017obr} 
  F.~Hagelstein and V.~Pascalutsa,
  Phys.\ Rev.\ Lett.\  {\bf 120}, no. 7, 072002 (2018).


\bibitem{Schwinger:1975ti} 
  J.~S.~Schwinger,
  Proc.\ Nat.\ Acad.\ Sci.\  {\bf 72}, 1 (1975); ibid.\ {\bf 72}, 1559 (1975)
  [Acta Phys.\ Austriaca Suppl.\  {\bf 14}, 471 (1975)].
  
\bibitem{HarunarRashid:1976qz} 
  A.~M.~Harun ar-Rashid,
  Nuovo Cim.\ A {\bf 33}, 447 (1976).

 \bibitem{Hagelstein2019} 
  F.~Hagelstein, V.~Pascalutsa and M.~Vanderhaeghen,
in preparation.


\bibitem{Pascalutsa:2012pr} 
  V.~Pascalutsa, V.~Pauk and M.~Vanderhaeghen,
  Phys.\ Rev.\ D {\bf 85}, 116001 (2012).




  
\bibitem{Knecht:2001qf} 
  M.~Knecht and A.~Nyffeler,
  Phys.\ Rev.\ D {\bf 65}, 073034 (2002).

\bibitem{Melnikov:2003xd} 
  K.~Melnikov and A.~Vainshtein,
  Phys.\ Rev.\ D {\bf 70}, 113006 (2004).

\bibitem{Hoferichter:2018dmo}
M.~Hoferichter, B.-L.~Hoid, B.~Kubis, S.~Leupold and S.~P.~Schneider, Phys.\ Rev.\ Lett.\ {\bf 121}, 112002 (2018).




\bibitem{Drell59}
S.~Drell, Il Nuovo Cimento {\bf 11}, 693-697 (1959).

\bibitem{Ametller:1984uk}
L.~Ametller, A.~Bramon, E.~Masso, Phys.\ Rev.\ D {\bf 30} 251 (1984).

\bibitem{Dorokhov:2007bd}
A.~E.~Dorokhov and M.~A.~Ivanov, Phys.\ Rev.\ D {\bf 75} 114007 (2007).

\bibitem{Nyffeler:2009tw}
A.~Nyffeler, Phys.\ Rev.\ D {\bf 79}, 073012 (2009).

\bibitem{Tanabashi:2018oca}
M.~Tanabashi, {\it et al.} (PDG), Phys.\ Rev.\ D {\bf 98}, 030001 (2018).

\bibitem{Griffiths}
D.~Griffiths, Introduction to Elementary Particles, WILEY-VCH Verlag, 2008.

  

\end{thebibliography}
\end{document}